\documentclass[pdflatex,sn-mathphys-num]{sn-jnl}
\usepackage{amsmath,amssymb,amsfonts}
\usepackage{graphicx}
\usepackage{booktabs}
\usepackage{float}
\usepackage{enumitem}

\title[Signal from Noise]{Signal from Noise: A Neural Network-Based Denoising Approach for Measuring Global Financial Spillovers}

\author*[1]{\fnm{Abdullah} \sur{Karasan}}\email{akarasan@umbc.edu}
\author[2]{\fnm{Özge Sezgin} \sur{Alp}}

\usepackage{booktabs}

\usepackage{graphicx}
\usepackage{float}
\usepackage{amsmath}
\usepackage{amssymb}

\begin{document}

\abstract{Filtering signal from noise is fundamental to accurately assessing spillover effects in financial markets. This study investigates denoised return and volatility spillovers across a diversified set of markets, spanning developed and developing economies as well as key asset classes, using a neural network-based denoising architecture. By applying denoising to the covariance matrices prior to spillover estimation, we disentangle signal from noise. Our analysis covers the period from late 2014 to mid-2025 and adopts both static and time-varying frameworks. The results reveal that developed markets predominantly serve as net transmitters of volatility spillovers under normal conditions, but often transition into net receivers during episodes of systemic stress, such as the Covid-19 pandemic. In contrast, developing markets display heightened instability in their spillover roles, frequently oscillating between transmitter and receiver positions. Denoising not only clarifies the dynamic and heterogeneous nature of spillover channels, but also sharpens the alignment between observed spillover patterns and known financial events. These findings highlight the necessity of denoising in spillover analysis for effective monitoring of systemic risk and market interconnectedness.}


\maketitle

\section{Introduction}
\label{sec:intro}

Interconnectedness is a pervasive phenomenon that must be considered to better understand the financial markets. As globalization and financial integration continue to advance, interconnectedness becomes increasingly important for investors, policymakers, and academics alike. Studying interconnectedness in the context of financial volatility enables us to understand the transmission of shocks, which is crucial for capturing the integration of financial risk. In particular, financial market volatility tends to increases and spreads across markets during times of crisis. Measuring such spillover effects is essential for providing early warning systems for emerging crises and for tracking the trajectory of ongoing crises \citep{diebold_yilmaz_2012}.

To this end, Diebold and Yılmaz (2009) introduced a spillover index to measure return and volatility spillovers based on variance decomposition of vector autoregressive models (VAR). Variance decompositions allows the forecast error variance to be apportioned among different sources within the system. Despite its usefulness, this method has certain limitations, most notably that the resulting variance decompositions are sensitive to the ordering of variables. Considering these drawbacks of the model, Diebold and Yilmaz (2012) (hereinafer, traditional spillover) proposed an improved spillover index based on the generalized VAR framework, which includes directional volatility spillovers and produces variance decompositions that are invariant to variable ordering.

The spillover index has been used in many studies to analyze the linkages between financial markets (e.g., \citep{zhou_2012,alter_2014,cronin_2014,corbet_2018,During_2021,li_2021,moratis_2021,reboredo_2021,choi_2022,wang_2022}). However, although there is a large literature addressing spillover effects, relatively few studies explicitly consider the role of noise in financial markets or seek to separate meaningful signals from random fluctuations \citep{hamdi_2019,jena_2022}. In general, there are two different effects on stock price dynamics: one originates from market participants making rational decisions, and the other from noise traders acting randomly. To understand the impact of rational strategies, it is important to remove the noise component from the series \citep{dremin_2008}. In detecting spillover effects across different markets, one should not ignore the extent to which the information is informative. Noise in the data affecting the utility of the information poses a challenge to the proper identification of spillover effects.

In this respect, the distinction between noise and signal not only helps us to better understand the volatility spillovers, but also allows us to identify spillovers caused by noise-aggregated, erratic information that does not lead to consistent and reliable results. Recent advances in deep learning, particularly the use of neural network-based denoising methods, offer a promising approach for extracting signals from noisy financial data. Effective denoising is increasingly vital, given the complexity of modern markets and the impact of noise trading on volatility and spillover effects \citep{jena_2022}. This paper adopts a feed-forward neural network to denoise covariance matrices prior to conducting spillover analysis, ensuring that the identified spillovers more accurately reflect genuine market interconnections rather than random noise.

Therefore, the first contribution of this study is to show the effect of denoising when examining the impact of spillovers across markets. By comparing results with and without neural network-based denoising, we provide direct evidence of the importance of noise removal in spillover analysis. Thus, return and volatility spillovers are analyzed both with and without denoising between different markets. This gives us valuable insights into the role of noise and approaches for addressing it, thereby enhancing our understanding of return and volatility spillovers.

The second contribution of the paper is to perform a return and volatility spillover analyses for different markets, i.e., developing and developed countries, to explain the possible variations in the spillover effects across countries with different levels of development. To this end, the impact of spillover effects is examined using the stock markets of both developed and developing countries.

The third contribution is to introduce denoised return and volatility spillovers measures in both static and time-varying frameworks. Given today's volatile world, it is more appropriate to develop spillover models that account for time-varying nature, as it is unlikely to propose a single and reliable model for the entire study period. Given the non-stationary and turbulent nature of modern financial markets, time-varying denoised spillover analysis powered by neural network-based filtering provides a robust and flexible tool for tracking and interpreting evolving market dynamics.

The remainder of this study is organized as follows: Section 2 provides a literature review, and Section 3 explains the denoised spillover methodology. Section 4 presents and discusses the empirical results, while Section 5 concludes the paper.

\section{Literature Review}
\label{sec:litreview}

Market spillover has important implications for asset management and global portfolio diversification, which has led to an ever-growing literature on global equity market interconnectedness. Zhou, Zhang, and Zhang (2012) analyzed directional volatility spillovers between the Chinese and global stock markets. They found that volatility in the Chinese market exerts a significant positive impact on other markets.

Li (2021) studied the time-varying volatility spillovers and the asymmetry of spillovers among the stock markets of the United States, Japan, Germany, the United Kingdom, France, Italy, Canada, China, India, and Brazil. The study concluded that global markets are highly interconnected and that volatility spillovers are time-varying, crisis-prone, and asymmetric. In addition, Li (2021) found that developed markets act as risk transmitters, while emerging markets act as risk receivers. Wang, Li, and Huang (2022) examined volatility spillovers and their time-varying dynamics among global stock markets with the five largest market capitalizations, including the stock markets of the United States, the United Kingdom, Japan, China, and Hong Kong, and took the S\&P 500, SZSE 300, Nikkei 225, Hang Seng, and FTSE 100. Accordingly, the stock markets of the United States and the United Kingdom are identified as net spillover senders, while the others are net receivers.

Abuzayed, Al-Fayoumi, and Jalkh (2021) studied the risk spillover effects among the global stock markets in the countries most affected by the pandemic COVID -19, including the U.S., Italy, Spain, Germany, China, France, the UK, Turkey, Switzerland, Belgium, the Netherlands, Canada, Austria, and Korea. During the COVID-19 period, they found that developed markets in Europe and North America both transmitted and absorbed more marginal extreme risk within the overall global market than Asian equity markets. Mensi et al. (2018) studied the volatility spillover between the global and regional equity markets of Greece, Ireland, Portugal, Spain, and Italy (GIPSI). They concluded that the U.S. global and regional markets are net senders, while the remaining equity markets are net receivers.

A series of global financial crises since the 1990s has prompted investors and portfolio managers to seek alternative strategies for portfolio diversification. Accordingly, commodities have also gained attention as financial assets suitable for portfolio diversification \citep{Al-Yahyaee_2019}. In addition, the large volume of currencies and stocks traded globally has led to speculation in these markets. Therefore, understanding the relationship between currencies and stock prices can also help fund managers to effectively manage risk \citep{kumar_2013}. For these reasons, a wide range of studies in the literature address the interconnectedness of different markets.

Kumar (2013) examined the return and volatility spillovers between exchange rates and stock prices of IBSA countries. The results suggest the existence of return and volatility spillover effects between markets. Al-Yahyaee et al (2019) analyzed risk spillovers among precious metals (silver, gold, palladium, platinum), energy commodities (crude oil, gasoline, fuel oil), and Saudi Arabia's stock markets. Their findings indicate a significant dynamic correlation between these different markets. Corbet et al. (2018) analyzed the relationships between cryptocurrencies and other financial assets and found a high degree of connectedness.

Hamdi et al. (2019) studied the volatility spillover effects between oil prices and sectoral indices in Gulf Cooperation Council (GCC) countries and concluded that all sectors are highly dependent on oil price volatility. Reboredo, Ugolini, and Hernandez (2021) examined spillovers among three market blocks: commodities, currencies, and equities. Their results suggest that equities spill over more to commodities, and commodities spill over more to currencies.

In recent years, deep learning approaches have emerged as powerful tools for analyzing and forecasting volatility spillovers, as well as for distinguishing between signal and noise in financial time series. Sahiner (2023) investigates volatility spillovers and contagion in major international stock-market crises spanning July 1997 to March 2021. The paper also develops an LSTM-based early warning system integrating DCC correlations to predict crisis onset. The model detected volatility spikes with high accuracy up to 12 months before both the Global Financial Crisis and COVID‑19 downturn, offering a robust signal for practitioners and policymakers. Karim, Shafiullah, and Naeem (2024) propose a novel methodology combining extreme value theory (EVT) with artificial neural networks to model extreme risk spillovers across 23 developed stock markets from January 1991 to July 2022. Their analysis reveals that extreme spillovers are significantly shaped by trade integration and economic interconnectedness, and tend to recur during prolonged crisis periods, highlighting Hong Kong and Western economies as pivotal nodes. This study demonstrates that machine learning–based approaches offer enhanced insights for risk management, effectively quantifying tail-risk transmissions and enriching traditional spillover analyses.

Building on the growing role of deep learning in financial econometrics, recent research has increasingly focused on neural network–based denoising methods for financial time series. Feiler (2024) introduces a novel autoencoder-based denoising method designed to significantly enhance the signal-to-noise ratio in financial time-series. The model jointly trains multiple autoencoders on both a target variable and various contextual variables, enforcing agreement across reconstructions to isolate consistent signals. This strong denoising approach improves the quality of volatility and risk estimation and offers a promising pre-processing step for downstream tasks like spillover analysis, risk modeling, and portfolio optimization. Song, Baek, and Kim (2021) introduce padding-based Fourier transform denoising (P‑FTD) as a preprocessing step before applying deep learning models (RNN, LSTM, GRU) to forecast stock indices such as the S\&P 500, SSE, and KOSPI. Their method removes noise by transforming the series into the frequency domain, zeroing out high-frequency components, and using symmetric padding to prevent endpoint distortion, before decoding back to the time domain. Results show that P‑FTD–enhanced models outperform their baseline counterparts in predictive accuracy and effectively mitigate time-lag issues associated with standard denoising techniques.

\section{Methodology}
\label{sec:methodology}
\subsection{Traditional Spillover Estimation}
In this study, the methodology of Diebold and Yilmaz (2012) is first applied to measure connectedness. Then, this methodology is enhanced by denoising the return and volatility matrix. In the remainder of this section, we attempt to convey how we modified Diebold and Yilmaz's (2012) spillover index through a denoising step, while summarizing the generalized VAR framework and structure of the spillover index.

In a covariance stationary N-variable VAR(p) model, each variable is expressed as follows:
\begin{equation}
x_t = \sum_{i=1}^{p}\phi_ix_{t-i} + \epsilon_t
\end{equation}
where $x_t$ is a Nx1 dimensional variable vector, $\epsilon_t \sim \mathcal{N}(0, \Sigma)$ is the i.i.d. disturbance term vector.

The moving average representation of the VAR(p) model is as follows:
\begin{equation}
x_t = \sum_{i=0}^{\infty}A_i\epsilon_{t-i}
\end{equation}
$A_i$ is defined recursively as $A_i = \sum_{j=1}^{p}A_{i-j}\phi_j, A_0 = I_N$ and $A_i = 0$ for $i<0$.

The H-step-ahead forecast-error variance decomposition can be written as follows:
\begin{equation}
\theta_{ij} ^g(H)=\frac{\sigma_{jj}^{-1}\sum_{h=0}^{H-1}(e_i^{'}A_h\Sigma e_j)^2}{\sum_{h=0}^{H-1}(e_i^{'}A_h\Sigma A_h' e_i)}
\end{equation}
Here, $\sigma_{jj}$ represents the standard deviation of the error term for the $j^{th}$ element, $A_h$ denotes the coefficient matrix of the moving average representation for the VAR(p) model, $\Sigma$ is the variance covariance matrix of the error term $\epsilon$, and $e_i$ denotes the selection vector which gets value one as the $i^{th}$ element and zeros otherwise. Return spillovers can similarly be derived by directly applying the generalized forecast-error variance decomposition to the returns themselves, rather than their residual covariance matrices. 

To compare pairwise spillovers each element of the variance decomposition matrix is normalized according to the row sum as follows:
\begin{equation}
\overline{\theta_{ij}^g}(H) = \frac{\theta_{ij}(H)}{\sum_{j=1}^{N}\theta_{ij}(H)}
\end{equation}
Then, $\sum_{j=1}^N \overline{\theta_{ij}^g}(H) = N$ and $\sum_{i,j=1}^N \overline{\theta_{ij}^g}(H) = 1$.

In line with the Diebold and Yilmaz (2012), total volatility spillover measuring the contribution of spillovers of volatility shocks accross variables to the total forecast error variance is defined as in equation \ref{total_spillover}.
\begin{equation}
S^g(H) = \frac{\sum_{\substack{i,j=1 \\ i\neq j}}^N\overline{\theta_{ij}^g}(H)}{\sum_{i,j=1}^N\overline{\theta_{ij}^g}(H)}* 100 = \frac{\sum_{\substack{i,j=1 \\ i\neq j}}^N\overline{\theta_{ij}^g}(H)}{N}* 100
\label{total_spillover}
\end{equation}

The directional volatility spillovers received by market i:
\begin{equation}
S_{i.}^g(H) = \frac{\sum_{\substack{j=1 \\ i\neq j}}^N\overline{\theta_{ij}^g}(H)}{\sum_{j=1}^N\overline{\theta_{ij}^g}(H)}* 100
\end{equation}

The directional volatility spillovers transmitted by market i:
\begin{equation}
S_{.i}^g(H) = \frac{\sum_{\substack{j=1 \\ i\neq j}}^N\overline{\theta_{ji}^g}(H)}{\sum_{j=1}^N\overline{\theta_{ij}^g}(H)}* 100
\end{equation}

Net volatility spillover from market i to all other markets:
\begin{equation}
S_i^g(H) = S_{.i}^g(H) - S_{i.}^g(H) 
\end{equation}

\subsection{Spillover Estimation with Neural Network Denoiser Architecture}
Traditionally, denoising of empirical covariance matrices $\mathbf{E}$ in high-dimensional finance relies on rotationally invariant shrinkage estimators \citep{bun_2016}. However, such methods have known limitations in capturing complex nonlinear dependencies and structural features, especially in the presence of market regimes or non-Gaussian behaviors.

In this study, we adopt a deep learning-based denoising approach to learn a data-driven mapping from noisy sample covariances to denoised, structure-preserving covariance matrices.

Let $\mathbf{S} \in \mathbb{R}^{N \times N}$ denote the sample covariance (or correlation) matrix of $N$ assets, computed from standardized returns in a rolling window. The denoising task is to learn a mapping
\[
\mathcal{D}_\theta: \mathbb{R}^{N \times N} \rightarrow \mathbb{R}^{N \times N}
\]
parameterized by neural network weights $\theta$, such that the output $\widehat{\mathbf{C}} = \mathcal{D}_\theta(\mathbf{S})$ provides a denoised estimator of the latent (true) covariance matrix $\mathbf{C}$.

The neural network denoiser proceeds as follows:
\begin{enumerate}
    \item Input Layer: The sample covariance matrix $\mathbf{S}$ is vectorized as $\mathbf{s} \in \mathbb{R}^{N^2}$.
    \item Feed-Forward Network: The input $\mathbf{s}$ is passed through several fully connected layers with nonlinear activations (e.g., GELU, LayerNorm):
    \begin{align*}
        \mathbf{h}_1 &= \mathrm{GELU}(\mathbf{W}_1 \mathbf{s} + \mathbf{b}_1) \\
        \mathbf{h}_2 &= \mathrm{LayerNorm}(\mathbf{W}_2 \mathbf{h}_1 + \mathbf{b}_2) \\
        &\vdots \\
        \mathbf{h}_{L-1} &= f_{L-1}(\mathbf{W}_{L-1} \mathbf{h}_{L-2} + \mathbf{b}_{L-1}) \\
        \mathbf{s}_{\text{den}} &= \mathbf{W}_L \mathbf{h}_{L-1} + \mathbf{b}_L
    \end{align*}
    \item Residual Connection and Symmetrization: The output vector is reshaped to $N \times N$ and symmetrized:
    \[
        \widehat{\mathbf{C}} = \alpha \mathbf{S} + (1 - \alpha) \cdot \mathrm{Sym}\left(\mathrm{Reshape}(\mathbf{s}_{\text{den}})\right)
    \]
    where $\alpha \in (0,1)$ is a residual weight (fixed or learned), and $\mathrm{Sym}(\mathbf{A}) = \frac{1}{2}(\mathbf{A} + \mathbf{A}^\top)$ ensures symmetry.
    \item Positive-Definiteness Correction: To ensure positive semi-definiteness, the eigenvalues of $\widehat{\mathbf{C}}$ are projected onto $\mathbb{R}^+$:
    \[
        \widehat{\mathbf{C}} = \mathbf{U} \, \mathrm{diag}\left( \max(\boldsymbol{\Lambda}, \epsilon) \right) \mathbf{U}^\top
    \]
    where $\mathbf{U}$ and $\boldsymbol{\Lambda}$ are the eigenvectors and eigenvalues of $\widehat{\mathbf{C}}$, and $\epsilon > 0$ is a small constant.
\end{enumerate}
\subsection{Loss Function}
The denoiser is trained by minimizing a structure-preserving loss function over a set of rolling-window covariance matrices:
\begin{equation}
    \mathcal{L}(\theta) = \frac{1}{B} \sum_{i=1}^B \left(
        \lambda_1 \left\| \mathcal{D}_\theta(\mathbf{S}_i) - \mathbf{S}_i \right\|_F^2
        + \lambda_2 \left\| \left[ \mathcal{D}_\theta(\mathbf{S}_i) \right]_{\mathrm{off}} - \left[\mathbf{S}_i \right]_{\mathrm{off}} \right\|_F^2
    \right)
\end{equation}
where $B$ is the batch size, $\|\cdot\|_F$ denotes the Frobenius norm, $\lambda_1, \lambda_2 > 0$ are weighting parameters, and $[\cdot]_{\mathrm{off}}$ denotes the off-diagonal elements. This loss penalizes both overall reconstruction error and off-diagonal (spillover-relevant) errors.

\subsection{Application to Spillover Analysis}
The output of the neural network-based denoiser architecture developed in Section 3.2 provides a denoised estimate of the residual covariance matrix, which we then use as a direct input to the volatility spillover analysis framework.

Recall from Section 3.2 that, given a sample covariance matrix $\hat{\boldsymbol{\Sigma}}$ (typically obtained from the residuals of a fitted VAR model), the neural network denoiser $\mathcal{D}_\theta$ is trained to produce a structure-preserving denoised matrix:
\begin{equation}
\hat{\boldsymbol{\Sigma}}^{(dn)} = \mathcal{D}_\theta(\hat{\boldsymbol{\Sigma}}).
\end{equation}
In this section, we integrate the denoised covariance matrix $\hat{\boldsymbol{\Sigma}}^{(dn)}$ into the standard spillover methodology. The spillover analysis is based on the variance decomposition of a vector autoregressive (VAR) model, following Diebold and Yilmaz (2012). Let $\mathbf{r}_t = (r_{1t}, r_{2t}, \dots, r_{N t})'$ denote the $N$-dimensional vector of asset returns at time $t$. The VAR($p$) model is:
\begin{equation}
\mathbf{r}_t = \sum_{k=1}^p \mathbf{A}_k \mathbf{r}_{t-k} + \boldsymbol{\varepsilon}_t,
\end{equation}
where $\mathbf{A}_k$ are the autoregressive coefficient matrices and $\boldsymbol{\varepsilon}_t$ is a zero-mean innovation with covariance $\boldsymbol{\Sigma}$.

Traditionally, the sample covariance matrix $\hat{\boldsymbol{\Sigma}}$ is used to compute the variance decompositions. In our approach, we replace this with the denoised version $\hat{\boldsymbol{\Sigma}}^{(dn)}$, directly utilizing the output from the neural network denoiser of Section 3.2. This substitution impacts every subsequent step of the spillover computation.

Specifically, we proceed to compute the moving average (MA) representation of the VAR model and the generalized variance decomposition (GVD) using the denoised covariance:
\begin{equation}
\mathbf{r}_t = \sum_{h=0}^{\infty} \boldsymbol{\Psi}_h \boldsymbol{\varepsilon}_{t-h},
\end{equation}
where $\boldsymbol{\Psi}_h$ are the MA coefficient matrices. The GVD at horizon $H$ is then calculated as:
\begin{equation}
\theta_{ij}^{(GVD, dn)}(H) = \frac{ [\sigma_{jj}^{(dn)}]^{-1} \sum_{h=0}^H \left( \mathbf{e}_i' \boldsymbol{\Psi}_h \hat{\boldsymbol{\Sigma}}^{(dn)} \mathbf{e}_j \right)^2 }
{\sum_{h=0}^H \left( \mathbf{e}_i' \boldsymbol{\Psi}_h \hat{\boldsymbol{\Sigma}}^{(dn)} \boldsymbol{\Psi}_h' \mathbf{e}_i \right) },
\end{equation}
where $\mathbf{e}_i$ is the $i$-th selection vector and $\sigma_{jj}^{(dn)}$ is the $j$-th diagonal element of $\hat{\boldsymbol{\Sigma}}^{(dn)}$.

Finally, the standard spillover measures (such as total, directional, and net spillovers) are constructed from the GVD matrix, following Diebold and Yilmaz (2012), but using the denoised covariance as input. This integration enables us to assess how deep learning-based denoising changes the estimated magnitude and structure of volatility spillovers across financial markets.
\section{Empirical Results}
\label{sec:results}

\subsection{Data}
The escalation of economic integration coupled with the increased frequency of cross-border transactions necessitates an examination of the interrelationships and contagion effects among various markets. Consequently, we have selected a cohort of ten equity markets, comprising five from developed nations and five from emerging economies. The developed nations included are Switzerland, Japan, the United States, the United Kingdom, and France, while the emerging economies encompass Turkey, Indonesia, Mexico, Iran, and India. Furthermore, we incorporate the cryptocurrency market, alongside commodities such as gold and oil, as well as foreign exchange markets, to investigate return and volatility spillovers across diverse asset classes.

The empirical investigation utilizes a daily dataset of significant global financial assets, covering the temporal span from January 1, 2014, to May 1, 2025. This dataset encompasses a wide array of instruments, including cryptocurrencies, commodities, foreign exchange rates, and equity indices sourced from both developed and emerging markets. An overview of the assets included in this analysis is provided in Table~\ref{tab:assets} in the Appendix. 

The data has been obtained from EOD Historical Data (EODHD), a reputable financial data provider, via their REST API. For each asset, the dataset comprises daily observations of the adjusted closing prices. The dataset is synchronized by date across all assets, thus ensuring temporal alignment for multivariate analysis. Rows containing missing data for all assets have been excluded to maintain data integrity.

To estimate the return and volatility spillovers, first, daily log returns are computed from adjusted closing prices. Then, volatility is computed over rolling 30-day standard deviation of daily returns. Thus, we obtain 2834 observations, and the descriptive statistics for asset returns and volatility can be found in Table~\ref{tab:descriptive_statistics_returns} and Table~\ref{tab:descriptive_stat_vol}, respectively.

\begin{table*}[htbp]
    \caption{Descriptive Statistics of Return.}
    \label{tab:descriptive_statistics_returns}
    \centering
    \small
    \setlength{\tabcolsep}{4pt}
    \begin{tabular}{lrrrrrrrr}
        \toprule
        & \textbf{Count} & \textbf{Mean} & \textbf{Std} & \textbf{Min} & \textbf{25\%} & \textbf{50\%} & \textbf{75\%} & \textbf{Max} \\
        \midrule
        BTC   & 2834 & 0.00268 & 0.04196 & -0.37170 & -0.01514 & 0.00153 & 0.02006 & 0.25247 \\
        BVSP  & 2834 & 0.00043 & 0.01481 & -0.14780 & -0.00704 & 0.00000 & 0.00807 & 0.13909 \\
        CUKX  & 2834 & 0.00027 & 0.00987 & -0.09588 & -0.00402 & 0.00028 & 0.00498 & 0.09421 \\
        ENX   & 2834 & 0.00097 & 0.01627 & -0.08413 & -0.00691 & 0.00073 & 0.00920 & 0.16331 \\
        EURUSD & 2834 & -0.00005 & 0.00553 & -0.03046 & -0.00311 & 0.00000 & 0.00295 & 0.03062 \\
        GLD   & 2834 & 0.00034 & 0.00889 & -0.05369 & -0.00445 & 0.00017 & 0.00496 & 0.04912 \\
        GSPC  & 2834 & 0.00043 & 0.01112 & -0.11984 & -0.00353 & 0.00030 & 0.00544 & 0.09515 \\
        JKSE  & 2834 & 0.00016 & 0.00936 & -0.07902 & -0.00422 & 0.00000 & 0.00485 & 0.08758 \\
        MXX   & 2834 & 0.00014 & 0.00960 & -0.06423 & -0.00494 & 0.00000 & 0.00536 & 0.04582 \\
        N225  & 2834 & 0.00039 & 0.01277 & -0.12396 & -0.00513 & 0.00005 & 0.00655 & 0.10226 \\
        NSEI  & 2834 & 0.00047 & 0.01003 & -0.12981 & -0.00392 & 0.00020 & 0.00560 & 0.08763 \\
        SSMI  & 2834 & 0.00016 & 0.00946 & -0.09637 & -0.00434 & 0.00030 & 0.00490 & 0.07016 \\
        USO   & 2834 & -0.00027 & 0.02401 & -0.25315 & -0.01149 & 0.00000 & 0.01233 & 0.16667 \\
        XU100 & 2834 & 0.00098 & 0.01532 & -0.09793 & -0.00658 & 0.00082 & 0.00925 & 0.09880 \\
        \bottomrule
    \end{tabular}
\end{table*}

Accordingly, Table~\ref{tab:descriptive_statistics_returns} shows that cryptocurrency (BTC) has the highest average return among all markets, with a mean value of approximately 0.27\%. In terms of risk, cryptocurrency again stands out with the highest standard deviation of returns (about 4.2\%), highlighting its notable price volatility. Among the equity indices, the French exchange (ENX), the Brazilian market (BVSP) and the Turkish market (XU100) show higher return variability.

Table~\ref{tab:descriptive_stat_vol} illustrates that cryptocurrency exhibits the highest mean volatility spillover (0.62) and the most greatest variability (0.25) among all examined assets, thereby underscoring its predominant function in the transmission of volatility across various markets. Within the realm of equity indices, France (ENX), Brazil, and Turkey (XU100) display comparatively high mean volatility figures, signifying a heightened responsiveness to global economic shocks. Conversely, EUR/USD and gold demonstrate lower average volatility, which is indicative of their relatively stable performance.

\begin{table*}[htbp]
    \caption{Descriptive Statistics of Volatility.}
    \label{tab:descriptive_stat_vol}
    \centering
    \small
    \setlength{\tabcolsep}{4pt}
    \begin{tabular}{lrrrrrrrr}
        \toprule
        & Count & Mean & Std & Min & 25\% & 50\% & 75\% & Max \\
        \midrule
        Cryptocurrency & 2820 & 0.61515 & 0.25020 & 0.14239 & 0.43904 & 0.58416 & 0.73398 & 1.52869 \\
        Brazil         & 2820 & 0.20978 & 0.10717 & 0.08866 & 0.15605 & 0.19207 & 0.23735 & 1.09415 \\
        UK             & 2820 & 0.13972 & 0.07134 & 0.06143 & 0.09708 & 0.12103 & 0.15441 & 0.60316 \\
        France         & 2820 & 0.23989 & 0.09692 & 0.07215 & 0.18218 & 0.22393 & 0.27571 & 0.85009 \\
        EURUSD         & 2820 & 0.08300 & 0.02908 & 0.03239 & 0.06170 & 0.07751 & 0.09686 & 0.20731 \\
        Gold           & 2820 & 0.13378 & 0.04244 & 0.06102 & 0.10364 & 0.12840 & 0.15376 & 0.35772 \\
        USA            & 2820 & 0.14780 & 0.09484 & 0.03567 & 0.09463 & 0.12330 & 0.17798 & 0.85310 \\
        Indonesia      & 2820 & 0.13364 & 0.06309 & 0.05194 & 0.09720 & 0.11986 & 0.14904 & 0.55432 \\
        Mexico         & 2820 & 0.14267 & 0.05186 & 0.05514 & 0.11153 & 0.13477 & 0.15882 & 0.45650 \\
        Japan          & 2820 & 0.18401 & 0.08267 & 0.06617 & 0.13458 & 0.16378 & 0.20785 & 0.56727 \\
        India          & 2820 & 0.14001 & 0.07555 & 0.05648 & 0.09982 & 0.12411 & 0.15746 & 0.74140 \\
        Switzerland    & 2820 & 0.13507 & 0.06485 & 0.05438 & 0.09678 & 0.11792 & 0.15130 & 0.53897 \\
        Oil            & 2820 & 0.33655 & 0.17604 & 0.10787 & 0.23213 & 0.30565 & 0.38163 & 1.47105 \\
        Turkey         & 2820 & 0.22726 & 0.08442 & 0.09292 & 0.17139 & 0.20576 & 0.25815 & 0.60395 \\
        \bottomrule
    \end{tabular}
\end{table*}

\subsection{Return Spillover}
This section presents the return spillovers and denoised return spillovers for all markets proposed by Diebold and Yilmaz (2012). Table~\ref{tab:returnspillover} shows the return spillover results for the entire sample period of 2014-01-01 and 2025-05-01.

The diagonal components in Table~\ref{tab:returnspillover}  denote the percentage of spillovers retained within each respective market, whereas the off-diagonal components illustrate the spillovers directed towards other markets. The columns labeled “FROM others” and “TO others” encapsulate the aggregate spillover that each market acquires from and conveys to other markets, respectively. Furthermore, the “NET” column emphasizes the net position of each market as either a transmitter or a receiver of return spillovers.

The findings reveal significant asymmetries in the transmission of return spillovers across markets. USA  (S\&P 500) emerges as the primary net transmitter, exhibiting a net value of 39.33, which underscores its considerable influence in disseminating shocks to other markets. Conversely, the Swiss market (SSMI) and Japan’s N225 present negative net values (-10.74 and -18.58, respectively), categorizing them as prominent net recipients of spillover effects.

Markets including the United Kingdom (CUKX, -4.29), Indonesia (JKSE, -4.49), and gold (GLD, -5.19) exhibit significant net receiver positions, predominantly functioning to absorb shocks rather than transmit them. In contrast, Cryptocurrency (BTC) and the oil demonstrate net values approaching zero (-0.09 and -0.35, respectively), indicating a comparatively balanced spillover profile with minimal impact as either transmitters or recipients.

Within the category of net senders, aside from GSPC, both EURUSD (6.37) and Mexico’s MXX (3.47) display positive net values, indicating their role as moderate sources of spillovers within the network. Conversely, other markets, including Brazil (-1.65), France (-1.50), India (NSEI, -0.68), and Turkey (XU100, -1.60), present negative net values, signifying a propensity to absorb shocks rather than transmit them. 

This net spillover distribution highlights the structural disparities inherent among global markets: A  few, particularly the US and EURUSD, serve as primary conduits for shock transmission, while a number of others, such as Switzerland, Japan, the UK, Indonesia, and gold, predominantly function as shock absorbers. Understanding these roles of transmitters and receivers is crucial for comprehending systemic risk, the pathways of contagion, and the interconnections within global financial markets.

Table~\ref{tab:denoisedreturnspillover} presents the denoised return spillover outcomes, providing a refined perspective on cross-market interactions by mitigating noise. The USA is reaffirmed as the predominant net transmitter, exhibiting a significantly elevated positive net value of 13.23, which underscores its pivotal role in disseminating return shocks to other markets. In contrast, the Swiss market (SSMI) is identified as the most substantial net receiver, with a value of 17.65, reflecting a pronounced inclination to absorb spillovers from the broader network. Additionally, Japan (-12.81), the United Kingdom (-9.03), Indonesia (-7.35), and gold (-4.65) continue to hold considerable positions as net receivers. 

The cryptocurrency market remains nearly neutral at -0.35, while Brazil (7.12) and EURUSD (9.51) are categorized as moderate net transmitters. Notably, several markets, including Mexico (-2.34) and Turkey (-2.58), demonstrate only minor net spillover values, indicative of a relatively balanced position within the global financial network.

A comparative analysis of traditional and denoised spillover estimates (Table~\ref{tab:returnspillover} and Table~\ref{tab:denoisedreturnspillover}) indicates that the denoising process not only diminishes the overall interconnectedness but also alters the hierarchy of net transmitters and receivers. For example, certain markets, including USA, continue to serve as principal transmitters, whereas others, such as Switzerland, Japan, the UK, and Indonesia, further entrench their positions as net receivers. Additionally, assets, specifically gold and oil, persist as net receivers in both traditional and denoised contexts, thereby reinforcing their status as safe-haven assets

Consequently, denoising clarifies which markets truly influence or absorb international return shocks, concurrently diminishing the perceived magnitude of total spillovers. This underscores the necessity for meticulous noise filtration in spillover analyses to accurately delineate the authentic channels of financial contagion and systemic risk across developed, emerging, and alternative markets.

\begin{table*}[htbp]
    \caption{Return Spillover Table}
    \centering
    \resizebox{\textwidth}{!}{
        \begin{tabular}{lrrrrrrrrrrrrrr|r|r}
            \toprule
            & \textbf{Cryptocurrency} & \textbf{Brazil} & \textbf{UK} & \textbf{France} & \textbf{EURUSD} & \textbf{Gold} & \textbf{USA} & \textbf{Indonesia} & \textbf{Mexico} & \textbf{Japan} & \textbf{India} & \textbf{Switzerland} & \textbf{Oil} & \textbf{Turkey} & \textbf{FROM} & \textbf{NET} \\
            \midrule
            Cryptocurrency & 98.54 & 0.24 & 0.25 & 0.05 & 0.04 & 0.01 & 0.20 & 0.05 & 0.06 & 0.06 & 0.31 & 0.04 & 0.13 & 0.02 & 1.46 & -0.09 \\
            Brazil & 0.09 & 95.57 & 0.27 & 0.40 & 0.59 & 0.06 & 0.43 & 0.28 & 0.96 & 0.48 & 0.57 & 0.06 & 0.09 & 0.16 & 4.43 & -1.65 \\
            UK & 0.03 & 0.36 & 90.30 & 0.23 & 0.55 & 0.29 & 4.03 & 0.10 & 1.54 & 0.28 & 1.88 & 0.33 & 0.08 & 0.01 & 9.70 & -4.29 \\
            France & 0.15 & 0.23 & 0.07 & 94.40 & 0.14 & 0.61 & 2.06 & 0.03 & 0.63 & 0.54 & 0.92 & 0.13 & 0.07 & 0.03 & 5.60 & -1.50 \\
            EURUSD & 0.06 & 0.02 & 0.61 & 0.57 & 96.67 & 0.66 & 0.67 & 0.27 & 0.08 & 0.03 & 0.03 & 0.20 & 0.01 & 0.11 & 3.33 & 6.37 \\
            Gold & 0.05 & 0.52 & 0.27 & 0.25 & 6.34 & 91.74 & 0.04 & 0.03 & 0.12 & 0.09 & 0.20 & 0.18 & 0.06 & 0.11 & 8.26 & -5.19 \\
            USA & 0.07 & 0.08 & 0.55 & 0.46 & 0.01 & 0.15 & 95.47 & 0.59 & 0.57 & 0.93 & 0.96 & 0.08 & 0.03 & 0.05 & 4.53 & 39.33 \\
            Indonesia & 0.11 & 0.51 & 0.28 & 0.41 & 0.22 & 0.25 & 2.69 & 93.25 & 0.75 & 0.51 & 0.18 & 0.24 & 0.29 & 0.31 & 6.75 & -4.49 \\
            Mexico & 0.12 & 0.27 & 0.15 & 0.64 & 0.23 & 0.11 & 0.21 & 0.31 & 96.44 & 0.46 & 0.24 & 0.58 & 0.16 & 0.08 & 3.56 & 3.47 \\
            Japan & 0.26 & 0.02 & 0.26 & 0.26 & 0.19 & 0.17 & 18.89 & 0.26 & 0.09 & 77.57 & 0.38 & 1.58 & 0.05 & 0.01 & 22.43 & -18.58 \\
            India & 0.06 & 0.21 & 0.42 & 0.66 & 0.04 & 0.43 & 5.24 & 0.15 & 0.85 & 0.26 & 90.79 & 0.48 & 0.36 & 0.05 & 9.21 & -0.68 \\
            Switzerland & 0.15 & 0.02 & 1.59 & 0.08 & 1.13 & 0.18 & 8.44 & 0.13 & 0.87 & 0.13 & 2.25 & 84.83 & 0.16 & 0.04 & 15.17 & -10.74 \\
            Oil & 0.19 & 0.16 & 0.02 & 0.04 & 0.13 & 0.01 & 0.68 & 0.02 & 0.38 & 0.00 & 0.10 & 0.14 & 98.09 & 0.03 & 1.91 & -0.35 \\
            Turkey & 0.02 & 0.14 & 0.67 & 0.04 & 0.12 & 0.13 & 0.27 & 0.04 & 0.12 & 0.08 & 0.53 & 0.39 & 0.05 & 97.40 & 2.60 & -1.60 \\
            \midrule
            TO others & 1.36 & 3.77 & 4.57 & 4.19 & 10.28 & 3.28 & 48.36 & 2.25 & 6.63 & 3.94 & 9.50 & 8.54 & 1.67 & 0.88 & 109.22 & \\
            \bottomrule
        \end{tabular}}
    \label{tab:returnspillover}
\end{table*}

\begin{table*}[htbp]
    \caption{Denoised Return Spillover Results.}
    \centering
    \resizebox{\textwidth}{!}{
        \begin{tabular}{l*{14}{r}|r|r}
            \toprule
            & \textbf{Cryptocurrency} & \textbf{Brazil} & \textbf{UK} & \textbf{France} & \textbf{EURUSD} & \textbf{Gold} & \textbf{USA} & \textbf{Indonesia} & \textbf{Mexico} & \textbf{Japan} & \textbf{India} & \textbf{Switzerland} & \textbf{Oil} & \textbf{Turkey} & \textbf{FROM} & \textbf{NET} \\
            \midrule
            Cryptocurrency      & 98.39 & 0.36 & 0.01 & 0.46 & 0.36 & 0.03 & 0.01 & 0.00 & 0.04 & 0.05 & 0.05 & 0.16 & 0.03 & 0.04 & 1.61  & $-0.35$ \\
            Brazil   & 0.07  & 92.91 & 0.10 & 0.69 & 0.41 & 0.07 & 2.20 & 0.00 & 0.13 & 0.08 & 0.41 & 2.75 & 0.12 & 0.06 & 7.09  & 7.12 \\
            UK   & 0.04  & 0.31 & 88.63 & 0.21 & 2.24 & 0.29 & 3.57 & 0.28 & 0.60 & 0.29 & 2.12 & 1.29 & 0.12 & 0.01 & 11.37 & $-9.03$ \\
            France      & 0.11  & 1.25 & 0.47 & 89.43 & 1.61 & 1.37 & 2.01 & 0.01 & 0.07 & 0.40 & 1.18 & 2.04 & 0.01 & 0.04 & 10.57 & $-2.62$ \\
            EURUSD & 0.16  & 0.29 & 0.23 & 0.06 & 97.38 & 0.31 & 0.55 & 0.01 & 0.06 & 0.03 & 0.08 & 0.82 & 0.00 & 0.02 & 2.62  & 9.51 \\
            Gold      & 0.12  & 0.43 & 0.03 & 1.24 & 2.85 & 92.79 & 0.38 & 0.16 & 0.07 & 0.03 & 0.39 & 1.22 & 0.15 & 0.14 & 7.21  & $-4.65$ \\
            USA   & 0.03  & 3.43 & 0.10 & 1.63 & 1.07 & 0.07 & 86.97 & 0.39 & 0.37 & 0.17 & 1.59 & 4.11 & 0.07 & 0.00 & 13.03 & 13.23 \\
            Indonesia   & 0.20  & 2.35 & 0.04 & 0.55 & 0.29 & 0.17 & 1.92 & 91.06 & 0.68 & 0.53 & 0.44 & 1.41 & 0.36 & 0.00 & 8.94  & $-7.35$ \\
            Mexico      & 0.04  & 0.91 & 0.04 & 1.01 & 0.38 & 0.11 & 1.38 & 0.42 & 94.02 & 0.06 & 0.48 & 1.08 & 0.03 & 0.04 & 5.98  & $-2.34$ \\
            Japan   & 0.04  & 0.13 & 0.06 & 0.13 & 0.58 & 0.01 & 7.47 & 0.02 & 0.02 & 85.15 & 0.68 & 5.42 & 0.19 & 0.10 & 14.85 & $-12.81$ \\
            India   & 0.03  & 2.27 & 0.67 & 0.08 & 0.04 & 0.05 & 4.05 & 0.19 & 1.24 & 0.09 & 89.81 & 1.43 & 0.02 & 0.03 & 10.19 & $-1.86$ \\
            Switzerland   & 0.23  & 0.50 & 0.05 & 0.93 & 1.91 & 0.04 & 2.09 & 0.00 & 0.22 & 0.23 & 0.73 & 92.91 & 0.10 & 0.07 & 7.09  & 17.65 \\
            Oil       & 0.02  & 1.52 & 0.07 & 0.10 & 0.28 & 0.03 & 0.35 & 0.00 & 0.03 & 0.03 & 0.07 & 2.63 & 94.86 & 0.00 & 5.14  & $-3.92$ \\
            Turkey  & 0.16  & 0.46 & 0.46 & 0.87 & 0.11 & 0.03 & 0.31 & 0.10 & 0.09 & 0.04 & 0.12 & 0.38 & 0.03 & 96.85 & 3.15  & $-2.58$ \\
            \midrule
            TO others & 1.25 & 14.22 & 2.33 & 7.96 & 12.13 & 2.57 & 26.27 & 1.59 & 3.63 & 2.04 & 8.33 & 24.75 & 1.22 & 0.57 & 108.85 & \\
            \bottomrule
        \end{tabular}}
    \label{tab:denoisedreturnspillover}
\end{table*}

In this section, we conduct a comparative analysis of traditional and denoised return spillovers over a specified period. It is important to note that in financial markets, the examination of data within a fixed temporal framework may obscure the effects of volatility arising from specific events, announcements, or crises. Consequently, significant dynamics instigated by such incidents may be disregarded. Therefore, an analysis of spillovers using a rolling window approach is essential.

\subsection{Time-Varying Traditional vs Denoised Return Spillover}
The investigation of time-varying spillovers provides critical insights, enabling the observation of fluctuations within the markets. Given the intricacies of contemporary financial systems, cyclical movements of this nature are frequently encountered. The ensuing results of this analysis pertain to the denoised time-varying return spillover.

\begin{figure}[H]
	\includegraphics[width=\textwidth]{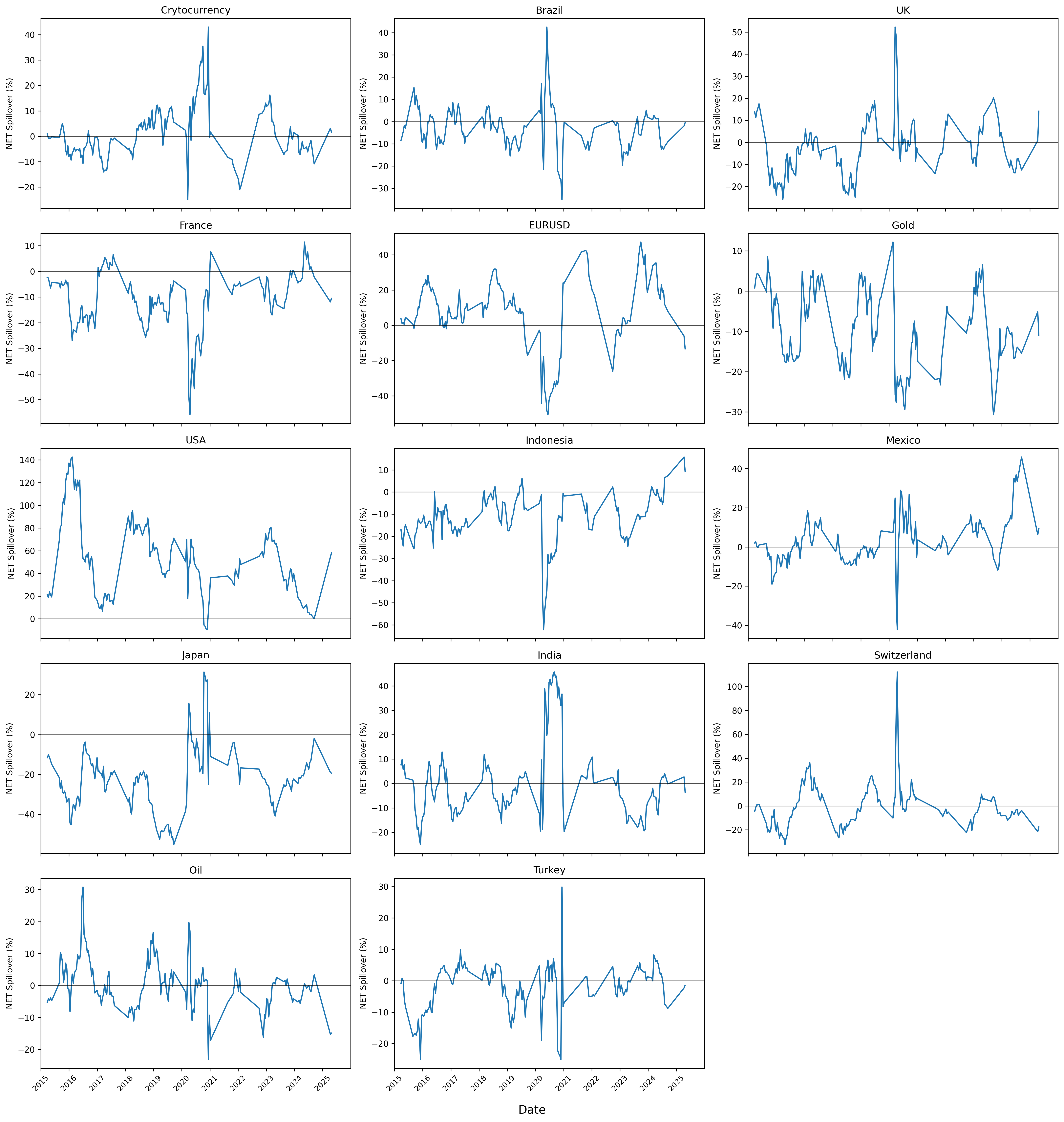}
	\caption{Time-Varying Traditional Return Spillover.}
	\label{timevaryingspilloverret}
\end{figure}

In Figure~\ref{timevaryingspilloverret}, the traditional rolling net return spillover findings are presented. The dynamics between net receivers and net transmitters across markets display significant instability and volatility, characterized by frequent fluctuations driven by noise that obscure fundamental trends. The markets for cryptocurrency, gold, and oil demonstrate unpredictable transitions between net receiver and net transmitter roles, with pronounced spikes and reversals notably intensifying during the year 2020. These alterations tend to be transient and rapidly reversed, suggesting a pronounced sensitivity to short-term noise rather than indicative of structural changes.

The exchange rate market is marked by persistent high oscillations, failing to uphold a consistent directional role as it alternates rapidly between net receiver and transmitter positions. Among national markets, developed economies such as Switzerland, and the United Kingdom display swift directional shifts, while emerging markets, including India, Turkey, and Brazil, this type of swift shifts are much more pronounced. Mexico, in particular, showcases pronounced return spillover volatility, with extreme fluctuations in both directions. This heightened return volatility in traditional spillover estimates indicates that markets respond acutely to short-term shocks and noise, complicating the identification of authentic structural changes instigated by significant events such as the Covid-19 pandemic. The traditional measure captures the erratic and unstable nature of financial market interconnectedness without differentiating between substantive signals and random variations.

\begin{figure}[H]
	\includegraphics[width=\textwidth]{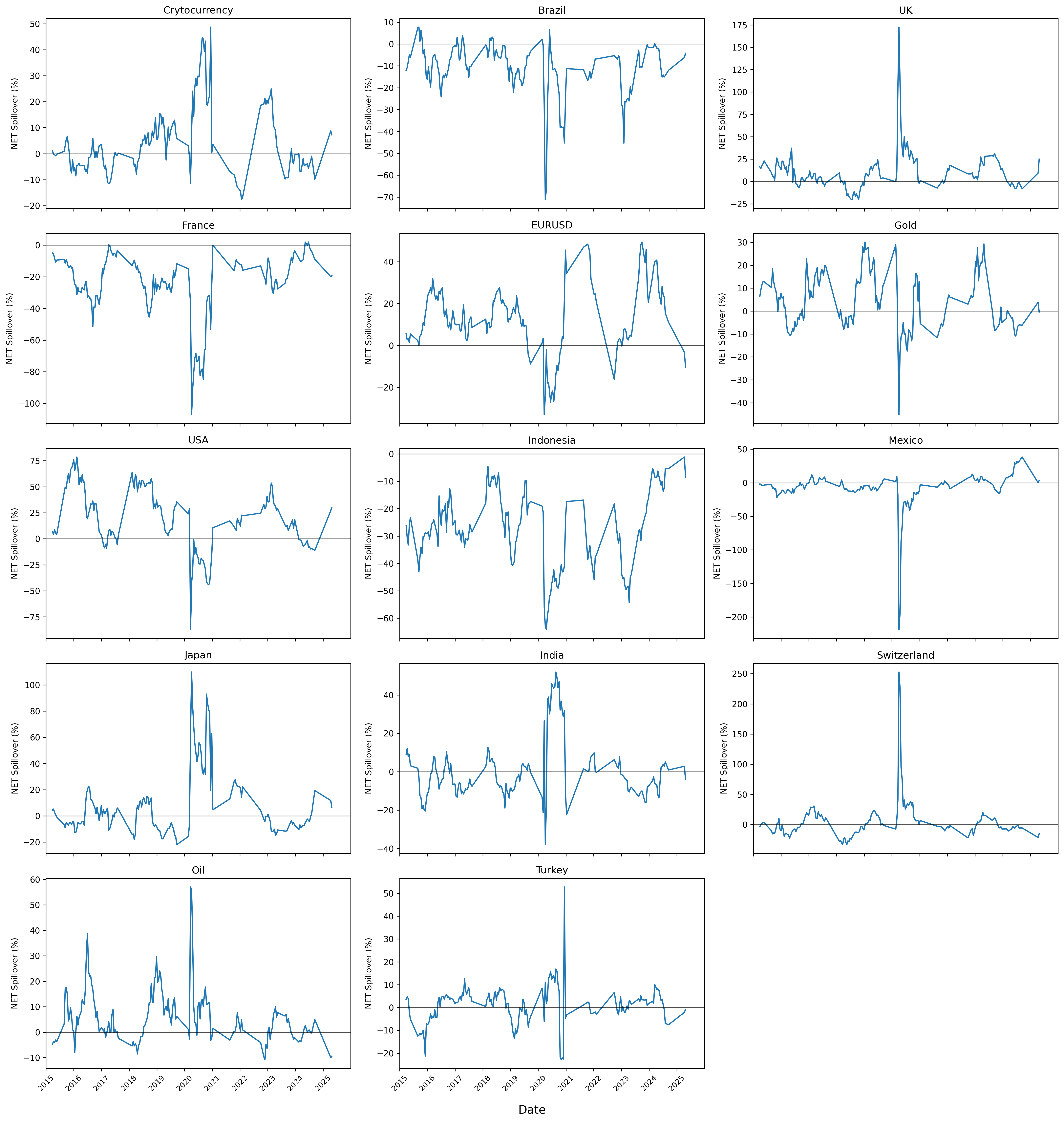}
	\caption{Time-Varying Denoised Return Spillover.}
	\label{timevaryingdenoisedspilloverret}
\end{figure}

The denoised return spillover results illustrated in Figure~\ref{timevaryingdenoisedspilloverret} reveal clearer patterns due to the reduction of noise, thereby highlighting more stable structural relationships. A majority of markets exhibit notable alterations in their net transmitter and receiver roles coinciding with the onset of the Covid-19 pandemic in early 2020, with these transformations demonstrating increased persistence and significance.

Cryptocurrency exhibits a distinct transformation from a net receiver to a net transmitter during the pandemic phase, whereas oil experiences a significant alteration to net receiver status amidst the crisis, particularly illustrated by the pronounced negative decline in 2020. The foreign exchange market retains a predominantly stable position as a net transmitter for the majority of the examined period, with a marked intensification of this function during times of crisis.

In the context of national markets, the United States exhibits a pronounced tendency towards net transmitter behavior, whereas Mexico reveals distinct intervals of transmission succeeded by phases of reception. India has experienced a significant transition from being a net receiver to emerging as a robust net transmitter during the pandemic, sustaining high levels of transmission in the subsequent period. In contrast, Indonesia and the United Kingdom demonstrate more stable trends relative to conventional metrics, with the United Kingdom exhibiting a consistent pattern of net transmitter behavior in recent times.

In consideration of the above discussion, the pandemic had widespread effects on how financial markets are connected, influencing spillover dynamics across a comprehensive range of economic development stages, from developed nations to emerging and developing economies, in significant and enduring manners. The refined findings indicate that no market category remained unaffected by the structural transformations instigated by the pandemic; rather, each group encountered specific yet substantial modifications in their positions within the global financial spillover network. In addition, even though the net receiver/transmitter roles are similar it is less pronounced when the spillover is denoised.

\subsection{Volatility Spillover}

Given these broader market conditions, it becomes increasingly important to understand how volatility shocks in one market propagate to others, a phenomenon known as volatility spillover. Analyzing volatility spillover dynamics provides critical insights into the interconnectedness of financial markets and helps identify both key transmitters and receivers of risk.

It is evident that denoising the volatility spillover matrix significantly reshapes the results, bringing out more nuanced and market-specific dynamics. As shown in Table~\ref{tab:StaticDenoisedVolatilitySpillover}, after denoising, the USA remains as the dominant net volatility transmitter with a net value of 30.45, followed by Oil (19.40) and EURUSD (18.29). Other notable transmitters include Mexico (13.01), and Cryptocurrency (6.71). In contrast, markets such as Brazil (–15.04), France (–15.10), India (–15.41), Indonesia (–14.71), UK (–10.53), Gold (–10.12), Switzerland (–4.31), Turkey (–2.27), and Japan (–0.37) are net receivers.

 Denoising further enhances specific spillover channels. For example, the total spillover from cryptocurrency to other markets is quantified at 10.76, an increase from the traditional spillover matrix value of 10.06, positioning it among the most significant transmitters following the United States (43.75) and Oil (24.78). The United States transmits 43.75 of its volatility to other markets, a substantial rise compared to the traditional findings of 36.43. Similarly, Oil experiences an increase from 22.88 in the traditional model to 24.78 in the denoised framework.
 
At the same time, denoising reveals that certain countries are more vulnerable to external shocks. France receives 20.26 of its volatility from others, with 0.74 coming specifically from cryptocurrency. Similarly, Brazil receives 23.95, India receives 21.37, and Switzerland receives 17.06 from other markets.

In comparison, the traditional spillover matrix (Table~\ref{tab:StaticVolatilitySpillover}) presents a different picture: the United States remains a net transmitter with a value of 20.63, while both Brazil (0.66) and Mexico (9.26) exhibit significantly lesser influence. The United Kingdom and India, identified as net receivers in the denoised framework (–10.53 and –15.41, respectively), display even greater net receiving characteristics in the traditional matrix, with values of –12.94 and –12.40. Notably, the aggregate spillover, indicated in the “TO others” row, demonstrates an increase post-denoising, with figures of 178.48 compared to 162.50, thereby underscoring the heightened interconnectedness illustrated by the denoised model.

Finally, denoising also shifts the direction of volatility transmission for some markets: for example, Mexico moves from being a moderate net transmitter (9.26 in traditional) to a stronger one (13.01), while Indonesia shifts from net transmitter (–9.11 in traditional) to net receiver (–14.71). The denoised average “FROM” value for all markets is 86.66, highlighting that after denoising, a significant proportion of volatility is still retained within markets, and only about 13\% is transmitted across markets.

\begin{table*}[htbp]
    \caption{Static Denoised Volatility Spillover Results.}
    \centering
    \resizebox{\textwidth}{!}{
        \small
        \begin{tabular}{l*{14}{r}|r|r}
            \toprule
            & \textbf{Cryptocurrency} & \textbf{Brazil} & \textbf{UK} & \textbf{France} & \textbf{EURUSD} & \textbf{Gold} & \textbf{USA} & \textbf{Indonesia} & \textbf{Mexico} & \textbf{Japan} & \textbf{India} & \textbf{Switzerland} & \textbf{Oil} & \textbf{Turkey} & \textbf{FROM} & \textbf{NET}  \\
            \midrule
Cryptocurrency    & 95.95 & 0.35 & 0.17 & 0.02 & 0.84 & 0.02 & 0.24 & 0.09 & 0.23 & 0.26 & 0.06 & 0.14 & 1.57 & 0.06 & 4.05 & 6.71 \\
Brazil   & 0.17 & 76.05 & 0.35 & 0.20 & 1.46 & 0.16 & 6.04 & 0.17 & 1.53 & 2.45 & 0.16 & 2.09 & 9.13 & 0.03 & 23.95 & -15.04 \\
UK   & 1.35 & 0.14 & 80.52 & 0.15 & 5.02 & 0.73 & 6.33 & 0.02 & 2.09 & 0.35 & 1.42 & 0.24 & 1.49 & 0.16 & 19.48 & -10.53 \\
France      & 0.74 & 1.70 & 2.50 & 79.74 & 3.25 & 2.69 & 3.03 & 0.45 & 0.48 & 0.93 & 1.84 & 0.94 & 1.31 & 0.38 & 20.26 & -15.10 \\
EURUSD & 1.68 & 0.64 & 0.06 & 0.02 & 96.39 & 0.25 & 0.09 & 0.09 & 0.08 & 0.02 & 0.28 & 0.16 & 0.15 & 0.08 & 3.61 & 18.29 \\
Gold      & 0.70 & 0.77 & 0.16 & 1.26 & 6.22 & 85.13 & 0.76 & 0.20 & 1.32 & 0.06 & 0.55 & 1.83 & 0.65 & 0.41 & 14.87 & -10.12 \\
USA   & 1.08 & 1.56 & 1.83 & 1.23 & 0.82 & 0.16 & 86.70 & 0.19 & 2.10 & 0.48 & 0.23 & 2.09 & 1.50 & 0.03 & 13.30 & 30.45 \\
Indonesia   & 0.08 & 0.95 & 0.86 & 0.45 & 0.29 & 0.13 & 6.70 & 82.66 & 4.27 & 2.27 & 0.31 & 1.00 & 0.04 & 0.01 & 17.34 & -14.71 \\
Mexico      & 0.09 & 0.18 & 0.20 & 0.57 & 0.49 & 0.19 & 1.47 & 0.04 & 95.96 & 0.41 & 0.12 & 0.08 & 0.15 & 0.06 & 4.04 & 13.01 \\
Japan   & 0.14 & 0.08 & 0.74 & 0.17 & 0.24 & 0.14 & 4.62 & 0.14 & 0.60 & 90.10 & 0.05 & 2.66 & 0.31 & 0.02 & 9.90 & -0.37 \\
India   & 0.67 & 2.39 & 0.38 & 0.19 & 0.09 & 0.08 & 6.39 & 0.76 & 2.20 & 1.50 & 78.63 & 0.14 & 6.47 & 0.12 & 21.37 & -15.41 \\
Switzerland   & 3.94 & 0.05 & 0.15 & 0.48 & 2.73 & 0.05 & 5.80 & 0.23 & 1.15 & 0.11 & 0.73 & 82.94 & 1.41 & 0.22 & 17.06 & -4.31 \\
Oil       & 0.07 & 0.09 & 0.61 & 0.07 & 0.35 & 0.08 & 1.87 & 0.17 & 0.54 & 0.32 & 0.16 & 1.01 & 94.62 & 0.02 & 5.38 & 19.40 \\
Turkey  & 0.04 & 0.01 & 0.94 & 0.35 & 0.09 & 0.07 & 0.41 & 0.09 & 0.46 & 0.36 & 0.05 & 0.38 & 0.61 & 96.15 & 3.85 & -2.27 \\
\midrule
TO others & 10.76 & 8.91 & 8.96 & 5.16 & 21.89 & 4.75 & 43.75 & 2.64 & 17.05 & 9.52 & 5.96 & 12.75 & 24.78 & 1.58 & 178.48 &  \\

            \bottomrule
        \end{tabular}}
    
    \label{tab:StaticDenoisedVolatilitySpillover}
\end{table*}

\begin{table*}[htbp]
    \caption{Static Traditional Volatility Spillover Results.}
    \centering
     \resizebox{\textwidth}{!}{
        \small
        \begin{tabular}{l*{14}{r}|r|r}
            \toprule
            & \textbf{Cryptocurrency} & \textbf{Brazil} & \textbf{UK} & \textbf{France} & \textbf{EURUSD} & \textbf{Gold} & \textbf{USA} & \textbf{Indonesia} & \textbf{Mexico} & \textbf{Japan} & \textbf{India} & \textbf{Switzerland} & \textbf{Oil} & \textbf{Turkey} & \textbf{FROM} & \textbf{NET}  \\
            \midrule
Cryptocurrency   & 95.293 & 0.707 & 0.171 & 0.037 & 0.743 & 0.042 & 0.239 & 0.135 & 0.228 & 0.200 & 0.075 & 0.138 & 1.935 & 0.056 & 4.707 & 5.356 \\
Brazil  & 0.096 & 85.843 & 0.189 & 0.170 & 0.716 & 0.147 & 3.283 & 0.146 & 0.840 & 1.033 & 0.123 & 1.172 & 6.225 & 0.018 & 14.157 & 0.661 \\
UK  & 1.354 & 0.291 & 79.508 & 0.227 & 4.490 & 1.250 & 6.282 & 0.039 & 2.095 & 0.268 & 1.955 & 0.246 & 1.848 & 0.146 & 20.492 & -12.942 \\
France   & 0.504 & 2.369 & 1.670 & 83.357 & 1.966 & 3.118 & 2.035 & 0.470 & 0.328 & 0.482 & 1.703 & 0.654 & 1.105 & 0.239 & 16.643 & -9.825 \\
EURUSD& 1.857 & 1.451 & 0.064 & 0.027 & 94.872 & 0.464 & 0.097 & 0.160 & 0.091 & 0.015 & 0.424 & 0.186 & 0.212 & 0.080 & 5.128 & 10.595 \\
Gold   & 0.431 & 0.978 & 0.098 & 1.203 & 3.436 & 90.006 & 0.464 & 0.188 & 0.815 & 0.030 & 0.462 & 1.158 & 0.498 & 0.234 & 9.994 & -3.499 \\
USA   & 1.057 & 3.153 & 1.770 & 1.866 & 0.721 & 0.276 & 84.200 & 0.289 & 2.063 & 0.360 & 0.304 & 2.096 & 1.821 & 0.025 & 15.800 & 20.625 \\
Indonesia  & 0.058 & 1.331 & 0.577 & 0.472 & 0.175 & 0.152 & 4.541 & 87.551 & 2.922 & 1.195 & 0.289 & 0.702 & 0.030 & 0.004 & 12.449 & -9.109 \\
Mexico   & 0.094 & 0.361 & 0.196 & 0.876 & 0.436 & 0.316 & 1.445 & 0.056 & 95.422 & 0.312 & 0.169 & 0.081 & 0.183 & 0.051 & 4.578 & 9.262 \\
Japan  & 0.171 & 0.209 & 0.916 & 0.325 & 0.269 & 0.292 & 5.756 & 0.265 & 0.756 & 87.028 & 0.090 & 3.418 & 0.478 & 0.026 & 12.972 & -7.629 \\
India  & 0.502 & 3.690 & 0.282 & 0.220 & 0.061 & 0.107 & 4.749 & 0.882 & 1.648 & 0.867 & 80.781 & 0.104 & 6.026 & 0.081 & 19.219 & -12.396 \\
Switzerland  & 3.841 & 0.108 & 0.145 & 0.725 & 2.373 & 0.083 & 5.597 & 0.340 & 1.125 & 0.085 & 0.973 & 82.695 & 1.711 & 0.199 & 17.305 & -6.097 \\
Oil   & 0.059 & 0.153 & 0.489 & 0.092 & 0.252 & 0.116 & 1.508 & 0.218 & 0.436 & 0.200 & 0.180 & 0.841 & 95.437 & 0.017 & 4.563 & 18.317 \\
Turkey & 0.038 & 0.017 & 0.984 & 0.578 & 0.086 & 0.133 & 0.430 & 0.152 & 0.491 & 0.295 & 0.074 & 0.412 & 0.807 & 95.504 & 4.496 & -3.320 \\
\midrule
TO others & 10.063 & 14.818 & 7.550 & 6.818 & 15.723 & 6.495 & 36.426 & 3.340 & 13.840 & 5.343 & 6.822 & 11.208 & 22.879 & 1.176 & 162.501 & \\
\bottomrule
        \end{tabular}}
    \label{tab:StaticVolatilitySpillover}
\end{table*}

\subsection{Time-Varying Traditional vs Denoised Volatility Spillover}
\begin{figure}[H]
	\centering
	\includegraphics[width=\textwidth]{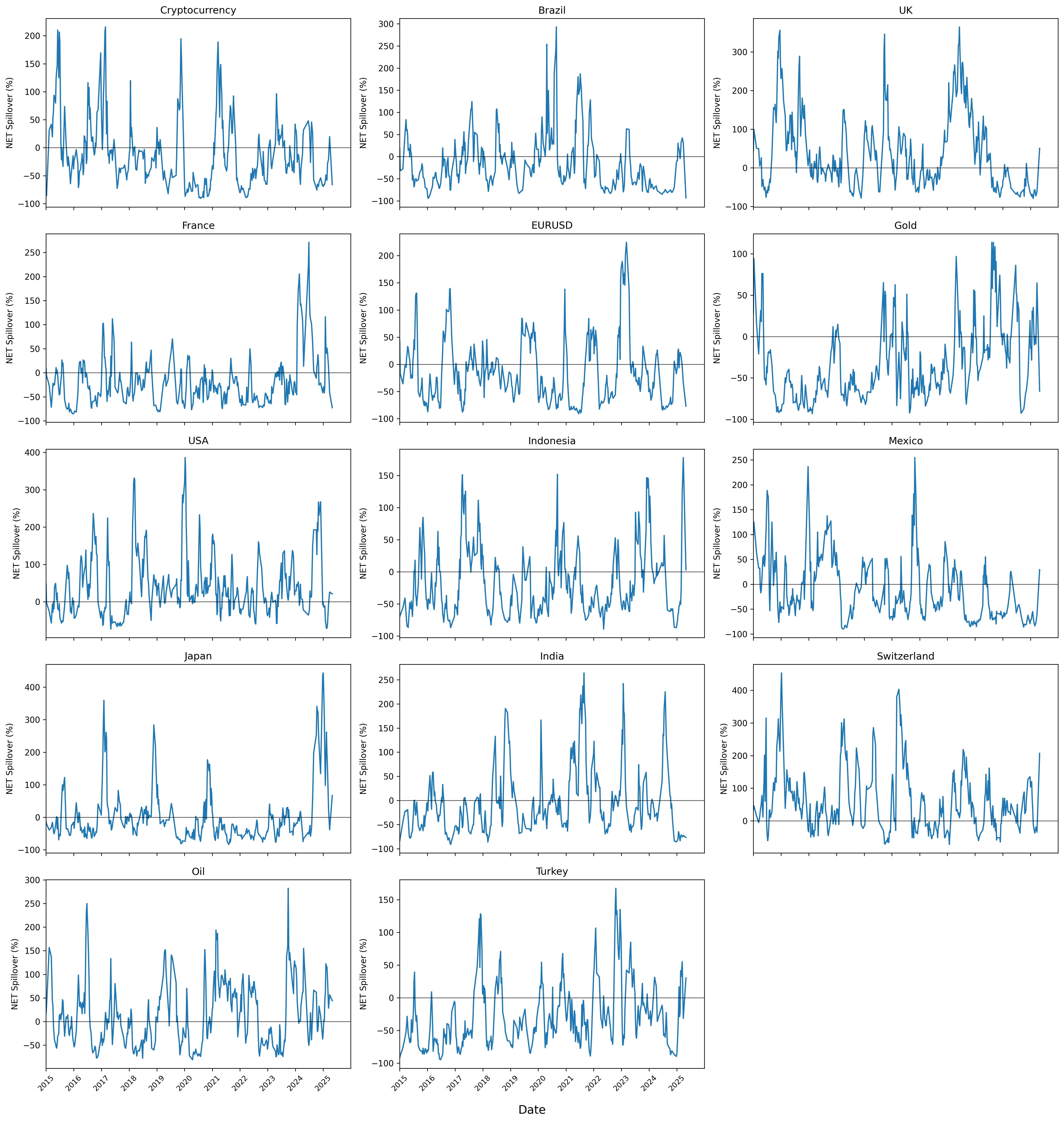}
	\caption{Time-Varying Traditional Volatility Spillover.}
	\label{rawtimevaryingspillovervol}
\end{figure}

It is essential to examine the temporal characteristics of traditional volatility spillovers to comprehend the manner in which financial interconnections react to external shocks and global occurrences. The dynamics of these volatility spillovers are depicted in Figure~\ref{rawtimevaryingspillovervol}. This figure shows the changing roles of transmission and reception among both developed and developing nations, as well as the assets involved over time.

The rolling net spillover analyses illustrated in Figure~\ref{rawtimevaryingspillovervol} indicate that both developed and developing markets, alongside commodities and currency markets, exhibit considerable variability in their functions as net transmitters or receivers of volatility over time. In the context of developed markets, including the United Kingdom, United States, Japan, Switzerland, and France, the net spillover indices do not demonstrate consistent characteristics of transmitters or receivers. Rather, their values exhibit significant fluctuations, characterized by notable spikes and frequent reversals. These variations are particularly pronounced during periods of systemic stress, especially around the year 2020, which aligns with the global disruption instigated by the Covid-19 pandemic. Such instances are marked by concurrent increases in net spillover across numerous developed markets, reflecting an elevated level of interconnectedness and swift shifts in the roles of volatility transmission.

Developing economies, such as Brazil, Turkey, Mexico, India, and Indonesia, demonstrate significantly greater volatility in their net spillover positions. The data representations for these countries reveal pronounced and frequent fluctuations, with rapid transitions between roles as transmitters and receivers. Notably, Turkey and Brazil experience intervals of substantial net transmission that are subsequently followed by sudden reversals, underscoring both an acute responsiveness to external disturbances and the inherent structural weaknesses within their evolving financial frameworks.

In both developed and emerging markets, there exists substantial evidence of synchronous behavior during global crises, as demonstrated by simultaneous increases in net spillover indices during significant stress events. Nevertheless, the magnitude and persistence of these spillover phenomena exhibit considerable variation. Certain markets exhibit short yet intense spikes, whereas others undergo extended and moderate variations.

In addition to equities, various asset classes, including cryptocurrency, gold, oil, and the EURUSD currency pair, exhibit dynamic net spillover characteristics. Notably, the cryptocurrency and oil markets experience pronounced volatility, particularly during episodes of global financial distress. Gold, traditionally regarded as a safe haven asset, reveals instances of both net transmission and reception, highlighting its changing function within the context of global risk dynamics. Furthermore, the EURUSD currency pair, being a significant financial instrument, mirrors global economic shocks, demonstrating considerable variations in net spillover effects.

Overall, Figure~\ref{rawtimevaryingspillovervol} highlights that neither market development status nor asset class guarantees a stable role as a volatility transmitter or receiver. Instead, net spillover positions are inherently time-varying, reflecting the complex interplay of systemic and idiosyncratic shocks. These findings reinforce the necessity for adopting a dynamic and adaptive framework when analyzing international volatility transmission and systemic risk, recognizing that all markets and asset classes are susceptible to pronounced shifts in their spillover behavior during periods of heightened uncertainty.

\begin{figure}[H]
	\centering
	\includegraphics[width=\textwidth]{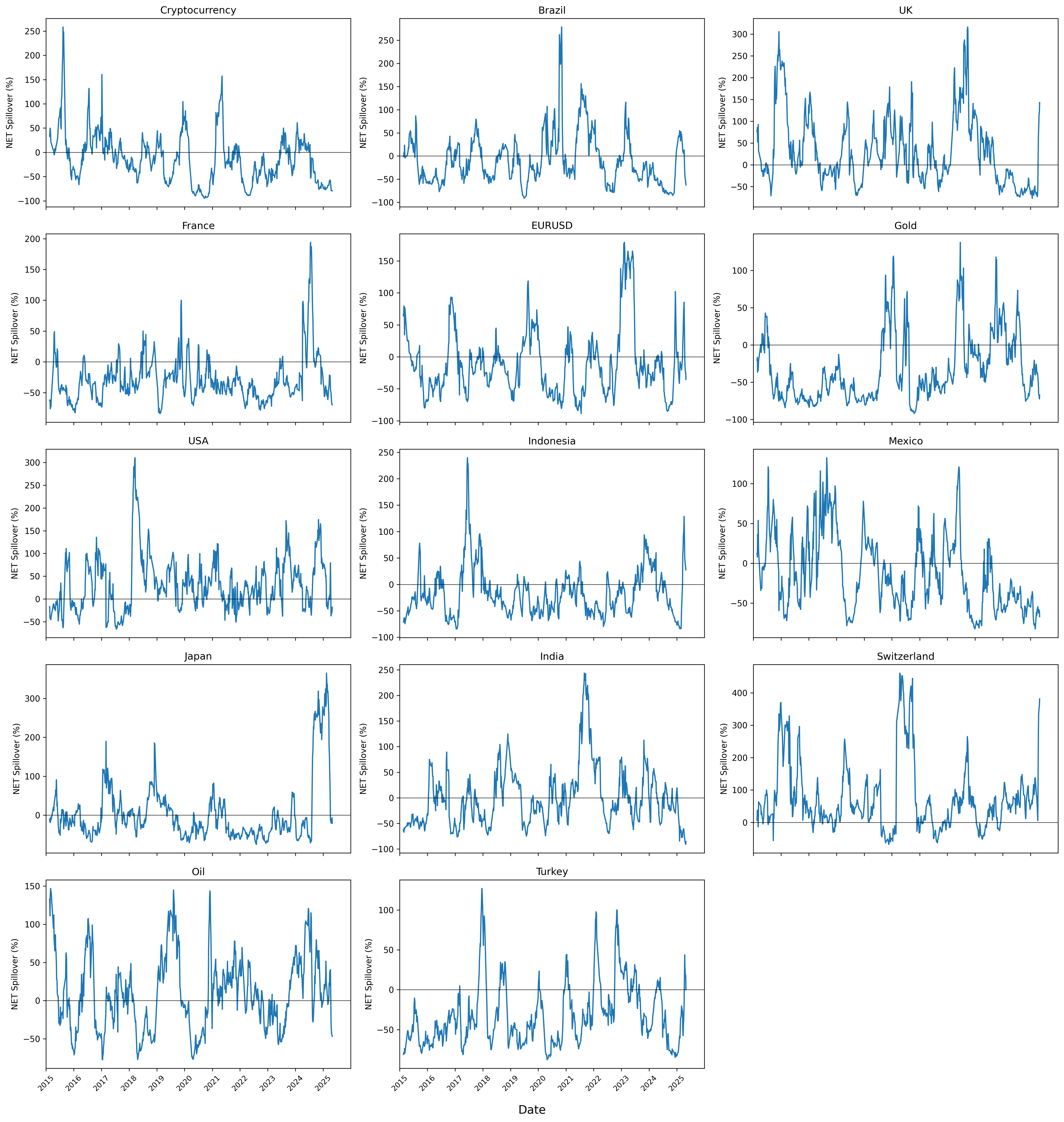}
	\caption{Time-Varying Denoised Volatility Spillover.}
	\label{denoisedtimevaryingspillovervol}
\end{figure}

As illustrated in Figure~\ref{denoisedtimevaryingspillovervol}, the time-varying characteristics of the denoised net volatility spillover offer a clearer and more interpretable perspective on the dynamics of global financial interconnectedness. Compared to traditional spillover estimates, the denoised series across all panels display smoother trends and more pronounced structural shifts, greatly reducing short-term noise and highlighting the evolution of each market’s role as a net transmitter or receiver of volatility.

Developed economies, including the United Kingdom, United States, Japan, France, and Switzerland, typically display more stable and interpretable net spillover patterns in the denoised visualizations. At the onset of the sample period, nations such as France and Japan are predominantly characterized as net receivers, whereas the United Kingdom and United States frequently act as net transmitters. This observation aligns with established narratives regarding the role of major financial centers in the propagation of global risk. Nonetheless, these roles are not immutable; during pivotal global events, especially at the commencement of the Covid-19 pandemic in 2020, a notable shift towards net receiver status is evident, highlighting the susceptibility of developed markets to significant systemic shocks. These regime transitions, which might be ambiguous or obscured by noise in conventional data, are rendered more discernible in the denoised series.

Emerging economies, such as Brazil, Turkey, Mexico, India, and Indonesia, frequently exhibit rapid and sometimes unexpected transitions between roles of transmitter and receiver. However, the application of denoising techniques serves to consolidate these fluctuations into a reduced number of significant regime changes. For example, Turkey illustrates distinct transitions as it initially acts as a receiver, subsequently becoming a transmitter for a duration, and then reverting to its original role, all of which correlate closely with recognized instances of domestic or regional market turmoil. Comparable patterns are observed in Brazil and Mexico, where pronounced variations in the conventional series are distilled into coherent and economically viable transitions following the denoising process.

Other asset classes such as cryptocurrency, gold, oil, and EURUSD also show refined and more coherent dynamics after denoising. Cryptocurrency and oil exhibit distinct spikes corresponding to periods of global financial stress. The EURUSD currency pair reflects similar shifts, confirming its sensitivity to broad-based risk transmission.

Importantly, the synchronization of markets during crises, exemplified by the Covid-19 pandemic, is significantly more pronounced in the denoised data series, where multiple markets tend to adopt net receiver status almost concurrently. This collective behavior, frequently obscured by short-term fluctuations in conventional spillover assessments, emerges as a salient characteristic of the denoised analysis. Nevertheless, the duration and magnitude of these transitions exhibit variability across different markets, indicative of regional resilience and the influence of policy interventions.

In summary, Figure~\ref{denoisedtimevaryingspillovervol} demonstrates that denoising enhances the practical relevance of net volatility spillover estimates. It enables researchers and policymakers to better distinguish between noise and substantive regime changes in volatility transmission, thus providing a sharper lens on the time-varying interdependencies that characterize the global financial system.

\section{Conclusion}
\label{sec:conclusion}
This study underscores the significant importance of denoising methodologies, particularly those based on neural networks, in enhancing the measurement and analysis of volatility spillovers within global financial markets. The refined results reveal that advanced economies, notably the United States, the United Kingdom, and Switzerland, frequently serve as principal conduits for risk transmission, aligning with their pivotal positions in the international financial architecture. Nonetheless, in times of severe market distress, exemplified by the Covid-19 pandemic, these markets collectively transition to net receiver roles, thereby exposing their susceptibility to external shocks. In contrast, emerging markets, represented by countries such as Turkey, Brazil, and Mexico, display more erratic and frequent shifts between transmitter and receiver statuses, indicating their heightened sensitivity to international contagion as well as the influence of domestic market dynamics. The denoising methodology effectively consolidates these volatile trends into more interpretable regime shifts and reveals instances of synchronization, particularly during global crises, when the directions of spillover effects temporarily align across various markets.

Furthermore, the process of denoising significantly diminishes noise and improves the alignment between spillover dynamics and actual events, such as financial crises or unique shocks. This enhancement in methodology not only bolsters the reliability of spillover analysis but also equips market participants and policymakers with more pertinent insights regarding the transmission of systemic risk and the interconnections within international markets. Ultimately, our findings underscore the necessity for any economic policy or risk management strategy that seeks to mitigate spillover effects to explicitly recognize the differentiation between signal and noise in financial data.
\newpage
\section{Appendix}


\begin{itemize}
\item \textbf{Funding:} This research received no external funding.
\item \textbf{Conflict of interest:} The authors declare no conflicts of interest.
\item \textbf{Data Availability:} Data can be provided upon request.
\end{itemize}

\begin{table*}[h!]
\centering
\begin{tabular}{@{}ll@{}}
\toprule
Abbreviation & Meaning \\
\midrule
VAR  & Vector Autoregressive \\
GVD  & Generalized Variance Decomposition \\
EVT  & Extreme Value Theory \\
LSTM & Long Short-Term Memory \\
DCC  & Dynamic Conditional Correlation \\
\bottomrule
\end{tabular}
\end{table*}

\begin{table*}[!htbp]
    \centering
    \caption{Summary of Assets Covered}
    \label{tab:assets}
    \begin{tabular}{lll}
        \toprule
        \textbf{Ticker}   & \textbf{Country/Region} & \textbf{Asset Class} \\
        \midrule
        BTC   & Global          & Cryptocurrency \\
        EURUSD & Eurozone/US     & FX Rate       \\
        GLD       & Global          & Gold          \\
        USO       & Global          & Oil           \\
        GSPC    & USA             & Equity Index (S\&P 500) \\
        CUKX     & United Kingdom  & Equity Index (FTSE 100) \\
        ENX       & France          & Equity Index (Euronext Paris) \\
        N225    & Japan           & Equity Index (Nikkei 225) \\
        SSMI    & Switzerland     & Equity Index (Swiss Market Index) \\
        NSEI    & India           & Equity Index (Nifty 50) \\
        BVSP    & Brazil          & Equity Index (Bovespa) \\
        MXX     & Mexico          & Equity Index (IPC) \\
        XU100   & Turkey          & Equity Index (BIST 100) \\
        JKSE    & Indonesia       & Equity Index (Jakarta Composite) \\
        \bottomrule
    \end{tabular}
\end{table*}


\end{document}